# Shear viscosity of the A$_1$- Phase of superfluid $^3He$

M. A. Shahzamanian   and   R. Afzali

Department of Physics, Faculty of Sciences,
University of Isfahan, 81744 Isfahan, IRAN

E-mail: shahzamanian @ hotmail.com

Abstract:

The scattering processes between the quasiparticles in spin- up superfluid with the quasiparticles in spin-down normal fluid are added to the other relevant scattering processes in the Boltzmann collision terms. The Boltzmann equation has been solved exactly for temperatures just below $T_{c1}$. The shear viscosity component of the $A_1$- phase drops as $C_1\left(1-\dfrac{T}{T_{c1}}\right)^{1/2}$. The numerical factor $C_1$ is in fairly good agreement with the experiments.

1. Introduction.

Viscosity of the $A_1$-phase of superfluid $^3$He is investigated at temperatures close to the transition temperature, $T_{c1}$, where the maximum gap in the excitation spectrum is small comparison with the thermal energy $K_B$ T. It is supposed that in the $A_1$-phase only spin-up pairs are existed, since it follows from the free energy expression that below the transition temperature, $T_{c1}$ it is lowered by formation of spin-up pairs. Formation of spin-down pairs becomes favorable below a temperature $T_{c2}$[1].

A number of viscosity measurements have been carried out in the $A_1$ and $A_2$ phases of superfluid $^3$He at different high magnetic fields[2,3]. The previous viscosity measurements were determined in low magnetic fields [4]. All the results show a sharp decrease proportional to the opening of the superfluid energy gap, $\Delta \propto \left(1 - \frac{T}{T_{c1}}\right)^{\frac{1}{2}}$ near $T_{c1}$ and goes as $T^{-2}$ for very low temperatures. The calculations of the former region is under construction and will be published elsewhere. In obtaining the values of viscosity coefficient on the whole region of temperatures is not analytically possible.

The viscosity of the A-phase, in zero magnetic field, has been calculated exactly for temperatures close to $T_c$ by Bhattacharyya et al (1977) [5] , and Pethick et al (1976) [6], and in the present of a mgnetic field has been considered by Shahzamanian (1975) [7]. The results of Bhattacharyya et al (1977) and Pethick et al (1976) on the viscosity drops as $\left(1 - \frac{T}{T_c}\right)^{\frac{1}{2}}$ for temperatures close to $T_c$ and the exact coefficient of $\left(1 - \frac{T}{T_c}\right)^{\frac{1}{2}}$ has been expressed as a function of normal state properties.

In this paper we use the Boltzmann equation approach for obtaining the viscosity of the $A_1$-phase for temperatures just below $T_{c1}$. One, therefore, needs to derive the correct form of the Boltzmann equation. This problem in the A-phase has been discussed by Bhattacharyya et al (1977) [5] and Pethick et al (1976) [6] extensively. The streaming terms in Boltzmann equation have the standard form , but the collision term is more complicated than normal-state. In a normal Fermi liquid at low temperatures the only important collision process is the scattering of pairs of quasiparticles, but in a superfluid the quasiparticle number is not conserved, so one also has to take into account decay processes in which a single quasiparticle decays into three, and the inverse processes, in which single quasiparticle coalesce to from one. In the $A_1$-phase one has to take into accunt other processes which come from the scattering between superfluid quasiparticles in the spin-up population, the so called Bogoliubov quasiparticles, and the normal fluid quasiparticles in the spin-down population. We shall evaluate the collision probabilities for these new processes.

The Boltzmann equation for a normal Fermi liquid has been solved exactly [8,9]. This equation has been also solved exactly for temperatures close to $T_c$ by Bhattacharyya et al (1977) [5]. The difference between the collision terms for the normal and superfluid states were treated as a perturbation. We use their method to solve the Boltzmann equation for the $A_1$-phase.

The paper is organized as follows. In section 2 by writing the interaction between the quasiparticles we obtain the collision terms for all the processes. Section 3 is allocated for writing the Boltzmann equation and its solution for the $A_1$-phase, then the shear viscosity tensor has been calculated for temperatures close to $T_{c1}$. Finally in section 4 we give some remarks and concluding results.

## 2. Collision integral.

For obtaining collision integral, we start with the interaction between the quasiparticles in the spin-up superfluid and spin-down normal fluid. This will be found by performing a Bogoliubov transformation on the normal-state interaction. The Bogoliubov transformation between the normal quasiparticle creation and annihilation operators $a^+_{\vec{p},\sigma}$ and $a_{\vec{p},\sigma}$ and the creation and annihilation operators $\alpha^+_{\vec{p},\sigma}$ and $\alpha_{\vec{p},\sigma}$ in the superfluid may be written as

$$a_{\vec{p},\sigma} = u(\vec{p})_{\sigma\sigma'}\alpha_{\vec{p},\sigma'} - \upsilon(\vec{p})_{\sigma\sigma'}\alpha^+_{-\vec{p},\sigma'}$$

$$a^+_{-\vec{p},\sigma} = \upsilon^*(\vec{p})_{\sigma\sigma'}\alpha_{\vec{p},\sigma'} + u(\vec{p})_{\sigma\sigma'}\alpha^+_{-\vec{p},\sigma'}$$

( 1 )

For the non-unitary state of the $A_1$-phase, we have the following properties between u and $\upsilon$ [10].

$$u(-\vec{P})_{\uparrow\uparrow} = u(\vec{P})_{\uparrow\uparrow}, \text{ and } \upsilon(-\vec{P})_{\uparrow\uparrow} = -\upsilon(\vec{P})_{\uparrow\uparrow}$$

( 2 )

The normal-state interaction is

$$H = \frac{1}{4} \sum_{1,2,3,4} \langle 3,4|T|1,2\rangle a_4^+ a_3^+ a_1 a_2 \qquad (3)$$

where i =1,2,3 and 4, stands for both momentum $(\vec{P}_i)$ and spin $(\sigma_i)$ variables. By using (1) in (3) the interaction between quasiparticles in the $A_1$-phase of superfluid is

$$H = \frac{1}{4} \sum_{\vec{P}_1,\vec{P}_2,\vec{P}_3,\vec{P}_4} \{T_t[-\upsilon^*(\vec{P}_4)_{\uparrow\uparrow}\alpha_{-\vec{P}_4\uparrow} - u(\vec{P}_4)_{\uparrow\uparrow}\alpha_{\vec{P}_4\uparrow}^+]$$

$$\times [-\upsilon^*(\vec{P}_3)_{\uparrow\uparrow}\alpha_{-\vec{P}_3\uparrow} - u(\vec{P}_3)_{\uparrow\uparrow}\alpha_{\vec{P}_3\uparrow}^+]$$

$$\times [u(\vec{P}_1)_{\uparrow\uparrow}\alpha_{\vec{P}_1\uparrow} - \upsilon(\vec{P}_1)_{\uparrow\uparrow}\alpha_{-\vec{P}_1\uparrow}^+]\;[u(\vec{P}_2)_{\uparrow\uparrow}\alpha_{\vec{P}_2\uparrow} - \upsilon(\vec{P}_2)_{\uparrow\uparrow}\alpha_{-\vec{P}_2\uparrow}^+] + \frac{1}{2}$$

$$\times (T_t - T_s)[-\upsilon^*(\vec{P}_4)_{\uparrow\uparrow}\alpha_{-\vec{P}_4\uparrow} - u(\vec{P}_4)_{\uparrow\uparrow}\alpha_{\vec{P}_4\uparrow}^+]\alpha_{\vec{P}_3\uparrow}^+[u(\vec{P}_1)_{\uparrow\uparrow}\alpha_{\vec{P}_1\uparrow} - \upsilon(\vec{P}_1)_{\uparrow\uparrow}\alpha_{-\vec{P}_1\uparrow}]$$

$$\times \alpha_{\vec{P}_2\uparrow} + \frac{1}{2}(T_t + T_s)\alpha_{\vec{P}_4\uparrow}^+[-\upsilon^*(\vec{P}_3)_{\uparrow\uparrow}\alpha_{-\vec{P}_3\uparrow} - u(\vec{P}_3)_{\uparrow\uparrow}\alpha_{\vec{P}_3\uparrow}^+]\;[u(\vec{P}_1)_{\uparrow\uparrow}\alpha_{\vec{P}_1\uparrow}$$

$$-\upsilon(\vec{P}_1)_{\uparrow\uparrow}\alpha_{-\vec{P}_1\uparrow}^+]\alpha_{\vec{P}_2\downarrow}\} \qquad (4)$$

where the amplitudes are given by

$$\langle\uparrow\uparrow|T|\uparrow\uparrow\rangle \equiv T_t, \langle\uparrow\downarrow|T|\uparrow\downarrow\rangle \equiv \frac{1}{2}(T_t + T_s) \quad and \quad \langle\uparrow\downarrow|T|\downarrow\uparrow\rangle \equiv \frac{1}{2}(T_t - T_s)$$

$$(5)$$

The gap parameter of the non-unitary state of the $A_1$- phase, $\Delta_{\vec{p}}$, has the same $\vec{P}$ dependence as the A-phase, i.e. it has the axial structure. Furthermore in the $A_1$- phase we may write $E_{\vec{p}}^2 = \varepsilon_{\vec{p}}^2 + |\Delta_{\vec{p}\uparrow\uparrow}|^2$, where $\varepsilon_{\vec{p}}$ is the normal-state quasipartile energy measured with respect to the chemical potential and $\Delta_{\vec{p}\uparrow\uparrow}$ is the magnitude of the gap in the direction $\vec{P}$ on the Fermi surface [10]. For the discussion of collision processes in the $A_1$-phase at temperatures just below the transition temperature it is most convenient to work in terms of quasiparticles which are related as closely as possible to the quasiparticle in the normal state. Accordingly we take the quasiparticle energy to be

$$E_{\vec{p}\uparrow} = \left(\varepsilon_{\vec{p}\uparrow}^2 + \Delta_{\vec{p}\uparrow\uparrow}^2\right)^{1/2} \operatorname{sgn}\varepsilon_{\vec{p}} \qquad (6)$$

Since we are interested only in changes in the collision integral of order $\Delta_{\vec{p}}$, we need, therefore, to retain only terms involving no more than a single $\upsilon_{\vec{p}}$ factor. As we have mentioned previously, in a superfluid the quasiparticle number is not conserved, and therefore scattering processes other than those in a normal Fermi liquid can occur. The first term in equation (4) indicates the scattering processes similar to those ones in the A-phase, i.e. two, decay and coalescence processes, which have been considered extensively by Bhattacharyya et al (1977) [5]. Here for brevity we write the final results of the collisions terms corresponding to the first term in equation (4)

$$(\frac{\partial n_1}{\partial t})_{coll.} = -\sum_{\vec{P}_2,\vec{P}_3,\vec{P}_4} \frac{2\pi}{\hbar} \frac{1}{4} |T_t|^2 n_1 n_2 (1-n_3)(1-n_4) \delta_{\vec{P}_1+\vec{P}_2,\vec{P}_3+\vec{P}_4} \delta(\varepsilon_1+\varepsilon_2-\varepsilon_3-\varepsilon_4)$$
$$\times [\Psi_1 + f_1(f_2\Psi_2 - f_3\Psi_3 - f_4\Psi_4)] \quad (7)$$

where $n_i$ is the quasiparticle distribution function, $\Psi_i$ is the deviation function defined in terms of the local equlibruim distribution function $n_i^{l.e.}(E_i)$ by the relation

$$n_i = n_i^{l.e.}[1-n_i^{l.e.}(E_i)]\Psi_i + n_i^{l.e.}(E_i) \quad (8)$$

and

$$f_{\vec{P}_i} = |u(\vec{P}_i)_{\uparrow\uparrow}|^2 - |\upsilon(\vec{P}_i)_{\uparrow\uparrow}|^2 = \frac{\varepsilon_{\vec{P}_i}}{E_{\vec{P}_i}} \equiv V_{\vec{P}_i} \quad (9)$$

For the $A_1$-phase we may write

$$f_{\vec{P}_i} = V_{\vec{P}_i} = |\varepsilon_{\vec{P}_i}|/(\varepsilon_{\vec{P}_i}^2 + \Delta_{\uparrow\uparrow}^2 \sin^2\theta)^{1/2} \quad (10)$$

where $\theta$ is the angle between $\hat{P}_i$ and the orbital anisotropy axis $\hat{l}$. The second and third terms in the interaction between the quasiparticles in equation (4) indicate the two quasiparticle scattering process and the coalensce scattering process between the Bogoliubov quasiparticles in the up-spin superfluid and the quasiparticles in the down-spin normal fluid. When the collision terms are linearized we have for the two quasiparticle scattering

$$(\frac{\partial n_1}{\partial t})_{coll} = -\sum_{\vec{P}_2,\vec{P}_3,\vec{P}_4} W'_{S,N}(1,2,3,4)[n_1 n_2(1-n_3)(1-n_4)(\Psi_1+\Psi_2-\Psi_3-\Psi_4)]$$

$$\times \delta_{\vec{P}_1+\vec{P}_2,\vec{P}_3+\vec{P}_4}\delta(\varepsilon_1+\varepsilon_2-\varepsilon_3-\varepsilon_4) \tag{11}$$

where

$$W'_{S,N}=\frac{2\pi}{\hbar}\frac{1}{4}[\frac{1}{4}|T_t-T_S|^2|u_1^2||u_4|^2+\frac{1}{4}|T_t+T_S|^2|u_1|^2|u_3|^2] \tag{12}$$

Note that we have replaced the superfluid quasiparticle energy by the corresponding normal-state energy in proper places, since this does not affect contribution of order $\Delta_{\vec{p}}$. For the process in which quasiparticle 1,2 and $-3$ coalesce to give quasiparticle 4 the linearized collision term is

$$(\frac{\partial n_1}{\partial t})_{coll}=-\sum_{\vec{P}_2,\vec{P}_3,\vec{P}_4}\frac{2\pi}{\hbar}\frac{1}{4}\{\,[\,|T_t-T_s|^2(|\upsilon_1|^2+|\upsilon_3|^2)\;](n_1n_2n_{-3}(1-n_4))$$

$$\times\delta(E_1+E_2+E_{-3}-E_4)(\Psi_1+\Psi_2+\Psi_{-3}-\Psi_4)+[|T_t+T_S|^2(|\upsilon_1|^2+|\upsilon_3|^2)]$$

$$\times(n_1n_2(1-n_3)n_{-4})\delta(E_1+E_2-E_3+E_{-4})(\Psi_1+\Psi_2-\Psi_3+\Psi_{-4})\}\,\delta_{\vec{P}_1+\vec{P}_2,\vec{P}_3+\vec{P}_4} \tag{13}$$

If we use the particle-hole symmetry of degenerate Fermi system we may replace $E_{-i}$ by $-E_i$, which has been used in obtaining equation (7) too. For the viscosity consideration, we may write $\Psi_2=\Psi_{-2}(-E_2)$. The collision integral (13), hence, may be rewrite as

$$(\frac{\partial n_1}{\partial t})_{coll}=-\sum_{\vec{P}_2,\vec{P}_3,\vec{P}_4}\frac{2\pi}{\hbar}\frac{1}{4}\frac{1}{4}\{[|T_t-T_S|^2(|\upsilon_4|^2+|\upsilon_1|^2)+|T_t+T_S|^2$$

$$\times(|\upsilon_3|^2+|\upsilon_1|^2)]n_1n_2(1-n_3)(1-n_4)(\Psi_1+\Psi_2-\Psi_3-\Psi_4)\}\delta_{\vec{P}_1+\vec{P}_2,\vec{P}_3+\vec{P}_4} \tag{14}$$

Finally by adding the collision terms in equations (11) and (14) we get

$$(\frac{\partial n_1}{\partial t})_{coll}=-\sum_{\vec{P}_2,\vec{P}_3,\vec{P}_4}\frac{2\pi}{\hbar}\frac{1}{4}n_1n_2(1-n_3)(1-n_4)\delta_{\vec{P}_1+\vec{P}_2,\vec{P}_3+\vec{P}_4}$$

$$\times\delta(\varepsilon_1+\varepsilon_2-\varepsilon_3-\varepsilon_4)\{[\frac{3}{2}|T_t|^2+\frac{1}{2}|T_S|^2]\Psi_1+[f_1f_2(\frac{3}{2}|T_t|^2+\frac{1}{2}|T_S|^2)$$

$$+\frac{1}{2}(1-f_1f_2)(|T_t|^2+|T_S|^2)]\Psi_2$$

$$-[f_1f_3(\frac{3}{2}|T_t|^2+\frac{1}{2}|T_S|^2)+\frac{1}{2}(1-f_1f_3)(|T_t|^2+|T_S|^2-\frac{1}{2}|T_t+T_S|^2)]\Psi_3$$

$$-[f_1 f_4 (\frac{3}{2}|T_t|^2 + \frac{1}{2}|T_s|^2) + \frac{1}{2}(1 - f_1 f_4)(|T_t|^2 + |T_s|^2 - \frac{1}{2}|T_t - T_s|^2)]\Psi_4\} \tag{15}$$

one obtains the normal-state collision integral by putting $f_i=1$ in equation (15).

## 3. Viscosity.

Before writing a formula for the shear viscosity we write the Boltzmann equation for the $A_1$-phase

$$-(\vec{P}_1)_i (\vec{v}_1)_j \frac{\partial n_1}{\partial E_1} (\frac{\partial u_k}{\partial r_l} + \frac{\partial u_l}{\partial r_k}) = (\frac{\partial n_1}{\partial t})_{coll} \quad . \tag{16}$$

where the collision integral operation is written in equation (15), and we pick out the terms in streaming terms of the Boltzmann equation which are relevant to viscosity. $u_k$ is the k component of a spatially varing velocity $\vec{u}$. One usually would like to express $\Psi_i$ in terms of the corresponding quantity for the normal-state. For this purpose it is more convenient to work with the function $X_i \equiv \Psi_i / V_i$, since in the superfluid close to $T_{c1}$ it differs from normal-state value by amounts of order

$\Delta_{\vec{p}} / K_B T_c$.

Hence equation (16) becomes

$$\hat{P}_{1i} \hat{P}_{1j} v_F P_F \frac{\partial n_1}{\partial \varepsilon_1} (\frac{\partial u_k}{\partial r_l} + \frac{\partial u_l}{\partial r_k}) = \sum_{\vec{P}_2, \vec{P}_3, \vec{P}_4} \frac{2\pi}{\hbar} \frac{1}{4} n_1 n_2 (1 - n_3)(1 - n_4)$$

$$\times \delta_{\vec{P}_1 + \vec{P}_2, \vec{P}_3 + \vec{P}_4} \delta(\varepsilon_1 + \varepsilon_2 - \varepsilon_3 - \varepsilon_4)$$

$$\{[\frac{3}{2}|T_t|^2 + \frac{1}{2}|T_s|^2]X_1 + [V_2^2 (\frac{3}{2}|T_t|^2 + \frac{1}{2}|T_s|^2) + \frac{1}{2}(1 - V_2^2)(|T_t|^2 + |T_s|^2)]$$

$$\times X_2 - [V_3^2 (\frac{3}{2}|T_t|^2 + \frac{1}{2}|T_s|^2) + \frac{1}{2}(1 - V_3^2)(|T_t|^2 + |T_s|^2 - \frac{1}{2}|T_t + T_s|^2)]X_3$$

$$-[V_4^2 (\frac{3}{2}|T_t|^2 + \frac{1}{2}|T_s|^2) + \frac{1}{2}(1 - V_4^2)(|T_t|^2 + |T_s|^2 - \frac{1}{2}|T_t - T_s|^2)]X_4\} \tag{17}$$

where we put $(\frac{V_i}{V_1} - V_i^2) \cong (1 - V_i^2)$, since it does not affect contribution of order $\Delta_{\vec{p}}$.

Now the Boltzmann equation may be replaced to a one-dimensional integral equation. For this purpose we define the function $Q(\hat{P}_1, t_1)$ as

$$\Psi_1 = \frac{\upsilon_F V_1}{K_B T} P_F \tau_o 2\cosh(\frac{t_1}{2})(\frac{\partial u_k}{\partial r_l} + \frac{\partial u_l}{\partial r_k})Q(\hat{P}_1,t_1) \qquad (18)$$

where $t = \frac{\varepsilon}{K_B T}$. By substituting equation (18) into equation (17) we get

$$\frac{\hat{P}_{1i}\hat{P}_{1j}}{\cosh(t/2)} = (\pi^2 + t^2)Q_{ij}(\hat{P}_1,t_1) - \int_{-\infty}^{\infty} dt' F(t-t')[\alpha_2 V^2(\hat{P}_1,t') \\ + \beta_2(1 - V^2(\hat{P}_1,t'))]Q_{ij}(\hat{P}_1,t') \qquad (19)$$

where $Q_{ij}(\hat{P},t) \equiv Q(\hat{P},t)\hat{P}_i\hat{P}_j$,

$$\alpha_2 = 2 < W_N(\theta,\phi)[-P_2(\cos\theta_{12}) + P_2(\cos\theta_{13}) + P_2(\cos\theta_{14})] > / < W_N(\theta,\phi) >, \qquad (20)$$

$$\beta_2 = 2 < \frac{2\pi}{\hbar}\frac{1}{4}[-\frac{1}{2}(|T_t|^2 + |T_S|^2)P_2(\cos\theta_{12}) + \frac{1}{2}(|T_t|^2 + |T_S|^2 - \frac{1}{2}|T_s + T_t|^2) \\ \times P_2(\cos\theta_{13}) + \frac{1}{2}(|T_t|^2 + |T_S|^2 - \frac{1}{2}|T_t - T_s|^2)P_2(\cos\theta_{14})] > / < W_N(\theta,\phi) >, \qquad (21)$$

$\theta_{ij}$ denotes the angle between $\vec{P}_i$ and $\vec{P}_j$,

$$< W_N(\theta,\phi) > \equiv \int \frac{d\Omega}{4\pi\cos(\theta/2)}\frac{2\pi}{\hbar}\frac{1}{8}(3|T_t|^2 + |T_S|^2) \qquad (22)$$

and

$$F(t-t') \equiv \frac{t-t'}{2\sinh[(t-t')/2]}$$

Bhattacharyya et al (1977)[5] by using the S- and P- wave approximation for the scattering amplitudes calculate the values of $\alpha_2$ and $<W_N(\theta,\phi)>$ for different values of pressure. Here we use Pfitzner (1985) procedure [11] for calculating the values of $\beta_2$ and $\alpha_2$ which appears in equation (19) . It should be noted that $\beta_2$ coefficient comes through the one-dimensional integral equation for the present of the new scattering processes in the A$_1$-phase. We can use quasiparticle scattering amplitude (QSA) of normal Fermi fluids instead superfluid QSA for temperatures near $T_{c\uparrow}$. By using general polynomial expansion of the QSA in normal Fermi fluids in Eqs. (20), (21) and (22) , namely [11] :

$$v(0)T_{s,t} = \sum_{k=0}^{\infty} \sum_{\ell=0}^{k} a_{\ell k} X_{\ell k}(v,P) \qquad (\ell \text{ even, odd}) \qquad (23)$$

where the coefficients with $\ell$ even (odd) belong to the singlet (triplet) part of the QSA,

$$X_{\ell k}(v,P) = (k+1)^{\frac{1}{2}} (2\ell+1)^{\frac{1}{2}} (P^2/4 - 1)^{\ell} P_{\ell}(v) P_{k-\ell}^{(2\ell+1,0)}(P^2/2 - 1)$$
$$k = 0,1,\ldots, \; ; \; \ell = 0,1,\ldots,k, \qquad (24)$$

$$P = 2\cos\frac{\theta}{2} \quad , \quad v = \cos\phi . \qquad (25)$$

And since we follow the procedure of Pfitzner (1985)[11] in calculating the QSA in the normal fluid at $T_{c\uparrow}$, we may truncate Eq.(23) at $k=3$ for pressure 34.4 bar. By using the values of $a_{\ell k}$ from table III of reference [11] and doing numerically the integrals in Eqs. (20) and (21), we get the following results:

$$\alpha_2 = 1.48 \quad and \quad \beta_2 = 0.72$$

Now we write a formula for the shear viscosity. The momentum flux tensor may be written as

$$\Pi_{lm} = \sum_{\vec{P}} \vec{P}_l (\frac{\partial E_{\vec{P}}}{\partial \vec{P}_m}) \delta n_{\vec{P}} \qquad (26)$$

where $\delta n_{\vec{P}} = n_{\vec{P}} - n_{\vec{P}}^{l.e.}$ characterize the deviation from local equlibrium, and from equation (8) we have

$$\delta n_{\vec{P}} = n_{\vec{P}}^{l.e.} (1 - n_{\vec{P}}^{l.e.}) \Psi_{\vec{P}} \qquad (27)$$

The shear viscosity is a fourth-rank tensor, which is defined by the relation

$$\Pi_{lm} = -\eta_{lmij} \left( \frac{\partial u_i}{\partial r_j} + \frac{\partial u_j}{\partial r_i} \right) \qquad (28)$$

In writing the above equation we have supposed that $l \neq m$ and $i \neq j$. When equation (27), with considering Eq.(18), is substituted in equation (26) and then compared with equation (28), we get

$$\eta_{lmij} = 15\eta << (VX_{lm}, VQ_{ij}) >> \equiv 15\eta Y_{lmij} \qquad (29)$$

where $\eta = \frac{1}{5}\rho \frac{m^*}{m} v_F^2 \tau_0$, $\tau_0$ is the characteristic relaxation time and is given by

$$\tau_0 = \frac{8\pi^4 \hbar^6}{m^{*3}(K_B T)^2 <W_N>} \quad , \tag{30}$$

$$X_{lm} \equiv \frac{\hat{P}_l \hat{P}_m}{\cosh(t/2)} \quad , \tag{31}$$

$$<<...>> = \int \frac{d\Omega_{\hat{P}}}{4\pi} ... \quad , \tag{32}$$

and

$$(A,B) = \int_{-\infty}^{\infty} A(t)B(t)dt \tag{33}$$

The kernel in the integral equation (19) has no structure on a scale $t \approx \frac{\Delta}{K_B T}$ and we may write equation (19) in the form

$$X_{ij} = (H_0 + H_1)Q_{ij} \tag{34}$$

where $H_0 Q_{ij}$ is the right-hand side of equation (19) with V=1, and

$$H_1 Q_{ij} \equiv \int_{-\infty}^{\infty} dt' F(t-t')(1-V^2(t'))(\alpha_2^2 - \beta_2^2)Q_{ij}(t')$$

$$= \pi(\alpha_2 - \beta_2)\widetilde{\Delta}_{\hat{P}} Q_{ij}(0) F(t) \tag{35}$$

In obtaining the last term in equation (35) we have used the following formula for any function A(t) having no structure on a scale $\Delta/K_B T$

$$\int_{-\infty}^{\infty} A(t)[1-V^2(t)] = A(0)\int_{-\infty}^{\infty}[1-V^2(t)]dt$$

$$= \pi A(0)\widetilde{\Delta} \tag{36}$$

where $\widetilde{\Delta} = \frac{\Delta}{K_B T}$. The dimensionless viscosity in equation (29) can be written as

$$Y_{lmij} = <<(X_{lm}, Q_{ij})>> - <<(X_{lm},(1-V^2)Q_{ij})>>$$

$$= <<(X_{lm}, Q_{ij})>> - \pi <<\widetilde{\Delta} X_{lm}(0), Q_{ij}(0)>> \tag{37}$$

we write $Q_{ij} = Q_{0ij} + Q_{1ij}$, where $Q_{0ij}$ is the unperturbed solution and $Q_{1ij} \propto \Delta_{\hat{P}}$ is the change due to the perturbation. By equating the terms independent of $\Delta_{\hat{P}}$ and those linear in $\Delta_{\hat{P}}$ to zero in equation (34), we have

$$X_{ij} = H_0 Q_{0ij}$$

$$0 = H_1 Q_{0ij} + H_0 Q_{1ij} \quad \text{or} \quad Q_{1ij} = -H_0^{-1} H_1 Q_{0ij} \tag{38}$$

where the first equation is the normal-state Boltzmann equation. Hence to the lowest order in $\widetilde{\Delta}$ we have

$$Y_{lmij} = \ll (X_{lm}, Q_{0ij}) \gg - \ll (X_{lm}, H_0^{-1} H_1 Q_{0ij}) \gg$$
$$- \pi \ll \widetilde{\Delta} X_{lm}(0), Q_{ij}(0) \gg \tag{39}$$

The second term in the right-hand side of equation (39) can be written as

$$\ll (X_{lm}, H_0^{-1} H_1 Q_{0ij}) \gg = \pi(\alpha_2 - \beta_2) \ll \widetilde{\Delta}_P Q_{0ij}(0) \int_{-\infty}^{\infty} dt\, F(t) Q_{0lm}(t) \gg \tag{40}$$

The integral term in equation (40) can be obtained simply by putting t = 0, V(t)=1 in equation (19). Finally we may write

$$\eta_{lmij} = \eta_{lmij}^n - 15\eta \ll \hat{P}_l \hat{P}_m \hat{P}_i \hat{P}_j \widetilde{\Delta}_{\hat{P}} \gg [\pi Q_N(0) \frac{\beta_2}{\alpha_2} + \pi^3 Q_N^2(0)(1 - \frac{\beta_2}{\partial_2})] \tag{41}$$

where $\eta_{lmij}^n = \eta(\delta_{li}\delta_{mj} + \delta_{lj}\delta_{mi})(X, Q_N) = \eta Y_N$. The values of $Q_N(0)$ and $Y_N$ has been evaluated by Bhattacharyya et al (1977) [5].

If the orbital axis is taken to be the Z axis, the shear viscosity has two different components, $\eta_{xy}$ and $\eta_{zx} = \eta_{zy}$. The angular expression in (41) for these components are

$$\ll P_x P_y P_x P_y \widetilde{\Delta}_{\hat{P}} \gg = (5\pi/256)\widetilde{\Delta}_{\max}$$

$$\ll P_z P_x P_z P_x \widetilde{\Delta}_{\hat{P}} \gg = (\frac{4}{5})(5\pi/256)\widetilde{\Delta}_{\max} \tag{42}$$

where $\Delta_{\max}$ is the maximum value of the $A_1$-phase gap parameter. Formula (41) and (42) give

$$\frac{\delta \eta_{xy}}{\eta_{xy}} = -(\frac{75\pi}{256})[\pi Q_N(o)\frac{\beta_2}{\alpha_2} + \pi^3 Q_N^2(o)(1 - \frac{\beta_2}{\alpha_2})]\widetilde{\Delta}_{\max}$$

and

$$\frac{\delta \eta_{zx}}{\eta_{zx}} = -(\frac{4}{5})(\frac{75\pi}{256})[\pi Q_N(o)\frac{\beta_2}{\alpha_2} + \pi^3 Q_N^2(o)(1 - \frac{\beta_2}{\alpha_2})]\widetilde{\Delta}_{\max} \tag{43}$$

## 4. Conclusions and some remarks.

To compare the results with the experiments [2,3,4] one has to know $\Delta_{max}$ as a function of temperature. Pethick et al (1976) [6] by taking the spin averaged gap generalize their results of the A-phase to the $A_1$-phase. In weak-coupling theory one has $\Delta_{max}(T) = (5/4)^{1/2} 3.06 K_B T_c (1 - \frac{T}{T_c})^{1/2}$ for the ABM-state and, hence, the spin averaged gap in the $A_1$-phase is $\Delta_{max} = 3.42[\frac{1}{2} K_B T_{C\uparrow}(1 - \frac{T}{T_{C\uparrow}})^{1/2}]$. By taking the strong coupling effect into account, finally we have $\Delta_{max} = 3.54[\frac{1}{2} K_B T_{C\uparrow}(1 - \frac{T}{T_{c\uparrow}})^{1/2}]$.

As we mentioned previously, these scattering processes between the quasiparticles in the up-spin superfluid and quasiparticles in the down-spin normal fluid play an important role in obtaining the Boltzmann equation for the $A_1$-phase. In this paper we take them into account and the results shows themselves through the factor $\beta_2$ in equations (19) and (41).

The values of $\alpha_2$ and $\beta_2$ depend slightly on the pressure through the Landau parameters. For pressures 21 bar and 34.36 bar, the melting pressure, the values of $\beta_2$ are respectively 0.79 and 0.72. The values of the last bracket in equation (43) for (21) and 34.36 bar pressures are respectively 2.27 and 2.12.

The viscosity data in the $A_1$ and $A_2$-phase of superfluid $^3$He were analyzed by Alvesola et al (1975)[4] in terms of a coefficient which gives the viscosity in the $A_1$-phase, and the result for temperatures close to $T_{c1}$ and at melting pressure is $\frac{\delta\eta}{\eta(T_{c1})} = -(2.7 \pm 0.2)(1 - \frac{T}{T_{c1}})^{1/2}$. This formula also fits with the data of Roobol et al(1994)[3]. Our results for P=21 bar are $\frac{\delta\eta_{xy}}{\eta_{xy}} = -2.09 \widetilde{\Delta}_{max}$ and $\frac{\delta\eta_{zx}}{\eta_{zx}} = -\frac{4}{5}(2.09)\widetilde{\Delta}_{max}$, and for the melting pressure we have $\frac{\delta\eta_{xy}}{\eta_{xy}} = -1.92 \widetilde{\Delta}_{max}$ and $\frac{\delta\eta_{zx}}{\eta_{zx}} = -\frac{4}{5}(1.92)\widetilde{\Delta}_{max}$. One expects that in the experiments the measured viscosity is $(\eta_{xy} + \eta_{zx})/2$ [6], Hence we have $\frac{\delta\eta}{\eta} = -3.30(1-\frac{T}{T_{c1}})^{\frac{1}{2}}, \frac{\delta\eta}{\eta} = -3.05(1-\frac{T}{T_{c1}})^{\frac{1}{2}}$ for pressures 21 and 34.36 bar respectively. We therefore see that agreement between our results and the experiments are fairly good.

Acknowledgment: This study has been financially supported by research council of the University of Isfahan, I.R.IRAN.